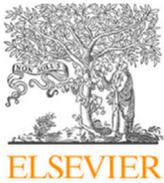
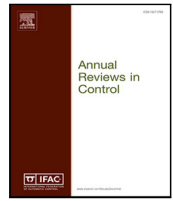

Full length article

# Lateral control for autonomous vehicles: A comparative evaluation

Antonio Artuñedo, Marcos Moreno-Gonzalez, Jorge Villagra *

*Centre for Automation and Robotics (CAR), CSIC - Universidad Politécnica de Madrid., Ctra. M300 Campo Real, km 0.200, Arganda del Rey, 28500, Madrid, Spain*

A B S T R A C T

The selection of an appropriate control strategy is essential for ensuring safe operation in autonomous driving. While numerous control strategies have been developed for specific driving scenarios, a comprehensive comparative assessment of their performance using the same tuning methodology is lacking in the literature. This paper addresses this gap by presenting a systematic evaluation of state-of-the-art model-free and model-based control strategies. The objective is to evaluate and contrast the performance of these controllers across a wide range of driving scenarios, reflecting the diverse needs of autonomous vehicles. To facilitate the comparative analysis, a comprehensive set of performance metrics is selected, encompassing accuracy, robustness, and comfort. The contributions of this research include the design of a systematic tuning methodology, the use of two novel metrics for stability and comfort comparisons and the evaluation through extensive simulations and real tests in an experimental instrumented vehicle over a wide range of trajectories.

## 1. Introduction

In the field of autonomous driving, the selection of an appropriate control strategy plays a key role in achieving the desired performance and ensuring safe operation. Numerous control strategies have been developed and employed to meet the various driving demands posed by different scenarios (Villagra, 2023). Understanding the strengths and weaknesses of these strategies is crucial to identify the most suitable controller for specific driving environments.

Although numerous path tracking controllers have undergone experimental tests, the existing literature lacks an objective and exhaustive comparative assessment of different lateral control structures. This assessment should encompass various lateral control structures and consider scenarios with a wide range of longitudinal and lateral velocities/accelerations. Advanced control formulations are often presented without being compared to finely-tuned simple controllers, while relatively simple control strategies emphasize their performance even under medium-high lateral acceleration levels (Sorniotti et al., 2016). Moreover, the subjective performance of different path tracking formulations, including the oscillation of control action and subsequent vehicle response, necessitates careful assessment through experimental tests. Such evaluations are crucial in order to obtain clear conclusions regarding the required level of sophistication in control systems.

This paper presents a comprehensive comparative evaluation of several prominent control strategies, namely Linear Quadratic Regulator (LQR), Model-Free Control (MFC), Speed-Adaptive Model-Free Control (SAMFC), Proportional–Integral–Derivative (PID), and Nonlinear Model Predictive Control (NLMPC). The primary objective is to evaluate and contrast the performance of these controllers across distinct driving purposes that capture a wide spectrum of real-world scenarios, reflecting the diverse needs and objectives of autonomous vehicles and driver assistance systems. To facilitate the comparative analysis, a comprehensive set of performance metrics, including IAE (Integral of Absolute Error), MLE (Maximum Lateral Error), and two additional metrics related to the frequency spectrum of the control action are employed. These metrics provide a multidimensional view of the controller performance, encompassing aspects such as accuracy, robustness and comfort.

Through a detailed examination of the experimental results and analysis of the data, this research aims to provide insights into the strengths and limitations of each controller strategy in relation to specific driving purposes. By comprehensively assessing the performance of these control strategies, researchers, engineers, and practitioners in the field of autonomous vehicles and driver assistance systems can make informed decisions when selecting an appropriate control strategy for a given driving scenario. The main contributions are summarized below:

- Two novel metrics are proposed to compare the behavior of controllers in terms of stability and comfort.

- The control strategies compared include both model-based and model-free state-of-the-art approaches that are tuned using the same optimization methodology.
- The evaluation is carried out from extensive simulation and real-world tests in an experimental instrumented vehicle.

The paper is structured as follows: Section 2 reviews state-of-the-art approaches for lateral control. Section 3 introduces the lateral dynamics with which the different controllers deal. In Section 4, the approaches implemented and evaluated in the comparison are described. Section 5 focuses on the evaluation methodology. This section introduces the benchmark, metrics and the tuning procedure of the tested controllers. An in-depth analysis of the experimental results is presented and discussed in Section 6. Finally, Section 7 draws some concluding remarks and future works.

## 2. Related work

In the field of autonomous driving, a large number of control strategies have been proposed to address the challenges posed by different driving scenarios (Arifin et al., 2019; Boyali et al., 2018; Liu et al., 2023). Several research works have focused on evaluating and comparing the performance of these control strategies, albeit with certain limitations.

Some of the proposed comparisons are limited to qualitative assessments. For example, the study proposed by Balaji and Srinivasan (2023) focuses on a qualitative analysis of merits and demerits of different controllers based on the results presented in different papers. It considers multi-constraint non-linear predictive controller, sliding mode control (SMC), neuro-fuzzy inference and two-point virtual control driving model, among others. An extensive qualitative analysis is also carried out in Kebbati et al. (2022), where authors focus on describing the strengths and weaknesses of 11 different control strategies including PID, Model-Predictive Control (MPC), SMC, LQR, $H_\infty$ among others. Lee et al. proposes a qualitative review of model-based and model-free control schemes (Rizk et al., 2023), remarking their advantages and disadvantages. In Samak et al. (2021), model-based and model-free are evaluated and compared in a qualitative manner. In Biswas et al. (2022), the authors compare PID and MPC controllers for lateral control from a general perspective, without detailing the limitations of each of the techniques.

Comparisons focused on quantitative assessment of different path tracking controllers are also found in the state of the art. Most of them are carried out in simulation. In Lee and Yim (2023) a comparison of PP, ST, LQR, PID, SMC and MPC methods in a low friction scenario is conducted in simulation. Abdallaoui et al. (2023) compare PID and MPC tracking quality, also in simulation. In Chaib et al. (2004), the authors focus on the evaluation of $H_\infty$, adaptive control, PID and Fuzzy techniques in a simulated environment. Moreover, in K. et al. (2019), Stanley, LQR and MPC Controllers are also compared in a simulation environment. Another comparison of tracking controllers is proposed in Calzolari et al. (2017), where the authors evaluate in simulation 8 different techniques that do not involve online optimization, including LQR, flatness-based control, and kinematic sliding mode, among others. In Menhour et al. (2017), a comparison of MFC, PID and non-linear flatness based control is carried out in an advanced simulation environment. Another comparison between LQR and MPC is performed in Yakub and Mori (2015), also limited to simulation. Finally, Stano et al. (2023) propose an extensive review of MPC approaches for path tracking ranging from methods using simple linear models with low degrees of freedom to complex nonlinear models. Most of the results provided in this review come from simulations, although there are also some results obtained from real vehicles.

Although in smaller numbers, comparisons are also found in experimental tests with real vehicles. In Dominguez et al. (2016), a kinematic controller based on the lateral speed is proposed and compared in terms of lateral error with pure pursuit (PP), Stanley (ST) and SMC. Pereira et al. (2023) proposes an adaptive reference aware MPC and is compared with other MPC-based techniques in terms of tracking quality both in simulation and real tests. In Chen et al. (2022), a MPC-based controller is proposed and compared with a feedforward-supported PID. The lateral deviation and computing time are analyzed in real experiments. A similar comparison is carried out in Hossain et al. (2022), where the proposed hybrid controller is compared against pure pursuit, LQR, Stanley and MPC.

The comparisons reviewed above have predominantly focused on limited scenarios with little variability. The restricted scope of these investigations raises concerns regarding the generalization of results in diverse and challenging real-world driving conditions. Most of the comparisons performed with real automated vehicles use data obtained from a single trajectory (Hossain et al., 2022; Pereira et al., 2023), and only (Dominguez et al., 2016) compares results from two paths. Given the dynamic and unpredictable nature of on-road scenarios, it becomes imperative to scrutinize the capabilities of path tracking controllers under a more extensive range of circumstances.

One of the key aspects when quantitatively comparing different control strategies is the methodology applied to fairly compare them. In this respect, it should be noted that most of the reviewed studies do not apply any common tuning methodology to the compared controllers and the comparative results presented are limited to one set of parameters for each controller. Those that do specify how the parameters were chosen are often imprecise, using definitions such as "well-tuned PID" (Samak et al., 2021) or, at best, using parameters provided in the corresponding cited papers, or set them manually when they are not provided (Calzolari et al., 2017). Other comparatives do not even specify the values of the controllers parameters used to obtain the results (Boyali et al., 2018; Hossain et al., 2022; K. et al., 2019). This makes it difficult to draw generalizable conclusions from the benchmarking results. Moreover, with regard to the metrics used, most of the reviewed studies focus on tracking quality, generally using lateral/angular error. When assessed, the stability, robustness and smoothness or comfort are often evaluated qualitatively from the control action or the error signal.

In summary, while the reviewed studies have provided valuable insights, they often are limited to qualitative analysis, lack uniform tuning methodologies, comprehensive analysis of limitations, or real-world experimental validation. Additionally, the selection of appropriate performance metrics to accurately capture the key aspects of controller performance remains a challenge. Most of the comparative studies found in the literature either are limited to qualitative analysis, or do not use a uniform and clear tuning methodology or are limited to simulation results. Thus, many authors presenting advanced control formulations tend to highlight their benefits without conducting comparisons with simpler controllers that have been fine-tuned. Conversely, authors presenting relatively simple control formulations tend to emphasize their performance in specific scenarios without providing results in a fair comparative framework. Furthermore, the subjective performance of different path tracking formulations, particularly in terms of control action oscillation and subsequent vehicle response, has not been thoroughly assessed. The limitations of previous comparative studies underscore the need for the present research, which seeks to address these gaps by conducting a comprehensive evaluation of path tracking controllers in a broader range of real-world scenarios, thus providing valuable insights into their applicability and performance across varied driving environments.

## 3. Lateral dynamics

Representing the vehicle dynamics with an accurate model is a complex task because real vehicles have (i) strong nonlinearities, e.g., road-tire friction, limits on the steering angle and gear shifting; (ii) dynamics that vary with longitudinal speed and steering angle; (iii) coupled lateral and longitudinal dynamics; and (iv) time-varying parameters, such as the road adherence, vehicle total mass and inertia.

These dynamics are often simplified by two prevalent models: the kinematic bicycle model and the dynamic bicycle model, also known as single-track model (see Arifin et al. (2019), Stano et al. (2023)). Even with the single-track simplification, couplings between lateral and longitudinal dynamics are present. The single-track model represents the vehicle lateral dynamics as follows:

$$\dot{v}_y = \frac{1}{m}(F_{x,f}\sin(\delta) + F_{y,f}\cos(\delta) + F_{y,r}) - v_x\dot{\Psi} \quad (1)$$

$$\ddot{\Psi} = \frac{1}{I_z}\left(l_f F_{x,f}\sin(\delta) + l_f F_{y,f}\cos(\delta) - l_r F_{y,r}\right) \quad (2)$$

where $m$ is the vehicle mass, $v_x$ and $v_y$ are the vehicle longitudinal and lateral speeds, $I_z$ is the yaw mass moment of inertia, $l_r$ and $l_f$ are the distances between the center of gravity (CoG) and the rear and front axle, respectively, $\dot{\Psi}$ is the yaw rate, $\delta$ is the steering angle and $F_{x,f(r)}$ and $F_{y,f(r)}$ are the longitudinal and lateral front (rear) tire forces, respectively.

The longitudinal dynamics are modeled as follows:

$$\dot{v}_x = \frac{1}{m}(F_{x,f}\cos(\delta) - F_{y,f}\sin(\delta) + F_{x,r}) + v_y\dot{\Psi} \quad (3)$$

Note that additional degrees of freedom (DoF) can be considered in terms of roll dynamics and vertical dynamics.

In order to simplify the synthesis of model-based controllers for automated vehicles, the system dynamics are typically decoupled into longitudinal motion and lateral motion. While the longitudinal vehicle behavior can be represented using a linear first order system model, the lateral behavior is inherently more intricate (Sorniotti et al., 2016). In the single-track model, the lateral forces are modeled as:

$$F_{y,f} = 2C_f\left(\delta - \arctan\left(\frac{v_y + \dot{\Psi}l_f}{v_x}\right)\right) \quad (4)$$

$$F_{y,r} = -2C_r \cdot \arctan\left(\frac{v_y - \Psi l_r}{v_x}\right) \quad (5)$$

where $C_f$ and $C_r$ are the cornering stiffness of the front and rear tires.

More detailed analytical tire models can be found in the literature, such as that used in Peterson et al. (2022); but the most precise are experimental models such as the one provided by Pacejka and Bakker (1992), which is significantly more complex than the single-track.

It is worth noting that the vehicle steering angle $\delta$ is not directly controlled; instead, it is handled through the steering torque with the following dynamics:

$$J_s\ddot{\delta}_d + B_u\dot{\delta}_d = T_c - T_{sa} \quad (6)$$

where $J_s$ is the moment of inertia of the vehicle's steering system, $B_u$ is the viscosity coefficient of the steering system, $T_c$ is the torque generated by the low-level steering actuator, $T_{sa}$ is the self-aligning torque and $\delta_d$ is the steering angle of the front wheel times the steering reduction coefficient of the vehicle.

The above described lateral vehicle dynamics are often simplified and linearly modeled as proposed in Rajamani (2011), which introduces a dynamic model in terms of error with respect to the road. To derive such linear model, some assumptions are made (Jiang & Astolfi, 2018): (i) small side slip is assumed, (ii) tires work on the linear region of the relationship between tire slip angle and the lateral force, (iii) the road-tire friction coefficient is assumed constant and (iv) the vehicle longitudinal speed is also assumed to be constant (Boyali et al., 2018). This model is commonly used in lateral control design (e.g., in Jiang and Astolfi (2018) and Zainal et al. (2017)) as it incorporates essential parameters to capture the vehicle's behavior accurately enough to keep a good balance with model complexity. The equations of the linearized single-track model in matrix form are in Eq. (7):

$$\begin{bmatrix}\dot{e}_y \\ \ddot{e}_y \\ \dot{e}_\Psi \\ \ddot{e}_\Psi\end{bmatrix} = \begin{bmatrix}0 & 1 & 0 & 0 \\ 0 & -\frac{2C_f+2C_r}{m v_x} & \frac{2C_f+2C_r}{m v_x} & \frac{-2l_f C_f+2l_r C_r}{m v_x} \\ 0 & 0 & 0 & 1 \\ 0 & -\frac{2l_f C_f-2l_r C_r}{I_z v_x} & \frac{2l_f C_f-2l_r C_r}{I_z v_x} & -\frac{2l_f^2 C_f+2l_r^2 C_r}{I_z v_x}\end{bmatrix}\begin{bmatrix}e_y \\ \dot{e}_y \\ e_\Psi \\ \dot{e}_\Psi\end{bmatrix} + \begin{bmatrix}0 \\ \frac{2C_f}{m} \\ 0 \\ \frac{2l_f C_f}{I_z}\end{bmatrix}\delta \quad (7)$$

where $e_y$ and $e_\Psi$ are the lateral and angular deviations of the vehicle from the path.

## 4. Implemented strategies

The control techniques evaluated in this work are LQR (Linear Quadratic Regulator), MFC (Model-Free Control), SAMFC (Speed-Adaptive Model-Free Control), PID (Proportional–Integral–Derivative), and NLMPC (Nonlinear Model Predictive Control). Their formulation is introduced in the following subsections, together with some relevant aspects of their implementation.

This techniques are selected in order to compare control strategies prevalent in the industry (PID) and typically used as comparison standard in academy (LQR) with modern model-based (MPC) and model-free (MFC, SAMFC) strategies.

### 4.1. LQR

Linear Quadratic Regulators (LQR) have been popularly applied for the vehicle lateral control. In Peterson et al. (2022) a LQR obtained with an extended vehicle model performs the lateral control in drifting maneuvers; an improved LQR strategy is applied in Wang et al. (2022) for the lateral control during a double lane changing maneuver; and a LQR with gain scheduling and power consumption considerations is applied in an electric vehicle in Han et al. (2018). The LQR controller implemented in this work is obtained with the vehicle's dynamic model from (7), which is discretized with a zero-order hold approximation and a sample time $T_s = 0.05$ s. The cost function is the usual in an infinite-horizon discrete-time LQR:

$$J = \sum_{k=0}^{\infty}\left(\mathbf{x}^T[k]Q\mathbf{x}[k] + u_{LQR}^T[k]Ru_{LQR}[k]\right) \quad (8)$$

where the state vector is $\mathbf{x} = \begin{bmatrix}e_y & \dot{e}_y & e_\Psi & \dot{e}_\Psi\end{bmatrix}^T$, the controlled variable $u_{LQR}$ is the steering angle $\delta$ and the matrices $Q$ and $R$ are defined as follows:

$$Q = \begin{bmatrix}q_1 & 0 & 0 & 0 \\ 0 & q_2 & 0 & 0 \\ 0 & 0 & q_3 & 0 \\ 0 & 0 & 0 & q_4\end{bmatrix}; \quad R = 1 \quad (9)$$

Note that the second and fourth states ($\dot{e}_y$ and $\dot{e}_\Psi$) are not directly measurable; however, they can be easily estimated from the first and third states respectively by applying a filtered derivative operator:

$$D(z) = \frac{1}{T_s}\frac{1 - z^{-1}}{N_{LQR} + (1 - C_1) \cdot z^{-1}} \quad (10)$$

As usual, the feedback matrix $K_{LQR} = \begin{bmatrix}k_1 & k_2 & k_3 & k_4\end{bmatrix}$ is obtained through the discrete-time algebraic Riccati equation (DARE). The control action is therefore defined as follows:

$$u_{LQR}[k] = K_{LQR} \cdot \mathbf{x}[k] \quad (11)$$

The LQR tunable parameters are $q_1$, $q_2$, $q_3$, $q_4$ and $N_{LQR}$.

*4.2. MFC*

Model-Free Control (Fliess & Join, 2013) is a recently developed control framework that has demonstrated its performance in a great variety of applications (Guilloteau et al., 2022; Moreno-Gonzalez et al., 2022; Villagra & Herrero-Perez, 2012; Villagra et al., 2020; Ziane et al., 2022). It is based on the reduction of the system's dynamics, that can be non-linear, time-varying or complex to identify, with a simple phenomenological model that is updated online, called *ultra-local* model:

$$y^{(n)} = F + \alpha \cdot u \quad (12)$$

in which the relationship between the input $u$ and the $n$th derivative of the output $y$ of the system is taken as linear with a constant ratio $\alpha$ that is a design parameter. This linear relationship is fitted by a variable $F$ that absorbs modeling errors and system disturbances.

The control loop is closed by a classical controller, being the intelligent PID (iPID) controllers the most usual:

$$u = \frac{1}{\alpha} \cdot \left( -F + y_r^{(n)} + K_p e + K_i \int e + K_d \dot{e} \right) \quad (13)$$

where $u$ is the control action, $y_r^{(n)}$ is the $n$th derivative of the output reference, $e$ is the tracking error and $K_p$, $K_i$ and $K_d$ are the control parameters, emulating those of a PID controller.

In the implementation, the term $F$ is updated continuously and must be estimated in real time by an estimator $\hat{F}$. A simple estimator is applied in the rest of the paper, it assumes $F$ to be constant between consecutive instants and thus can be estimated from previous control actions from (12) as follows:

$$\hat{F}(t_k) = \hat{y}^{(n)}(t_k) - \alpha \cdot u(t_{k-1}) \quad (14)$$

where $t_k$ is the current instant and $\hat{y}^{(n)}$ is the estimation of the $n$th derivative of $y$. This estimation is obtained by applying the following filtered derivative operator $n$ times to the output of the system:

$$D(z) = \frac{1}{T_s} \frac{1 - z^{-1}}{C + (1 - C) \cdot z^{-1}} \quad (15)$$

being $T_s$ the sample time and $C$ the filtering parameter, which is designed so that measurement noise is reduced.

The MFC controller implemented in this work is a second order iPD ($n = 2$). Therefore, (13) yields the control action $u_{iPD}$:

$$u_{iPD}(t_k) = \frac{-\hat{F}(t_k) + \ddot{y}_{1r}(t_k) + K_p e(t_k) + K_d \hat{\dot{e}}(t_k)}{\alpha} \quad (16)$$

$$e(t_k) = y_{1r}(t_k) - y_1(t_k); \quad \hat{\dot{e}}(t_k) = \dot{y}_{1r}(t_k) - \hat{\dot{y}}_1(t_k)$$

where $y_1$ is the lateral deviation of the vehicle, $e$ is the tracking error and $\hat{\dot{e}}$ is the filtered estimation of the tracking error derivative.

The filtering parameter is fixed by design to $C = 1.5$, so the tunable parameters are $\alpha, K_p, K_d$.

*4.3. SAMFC*

Speed-Adaptive Model-Free Control (Moreno-Gonzalez et al., 2022) is a variation of the original MFC structure that is specifically developed for the lateral control of autonomous vehicles. It has been shown in Moreno-Gonzalez et al. (2022, 2023) that the MFC control parameter $\alpha$ is related to the aggressiveness of the controller. In lateral vehicle control, when the longitudinal speed of the vehicle is high, aggressive MFC controllers may become unstable; on the contrary, when the speed is low, smooth controllers do not accurately follow the path. Therefore, the variation of $\alpha$ (~ aggressiveness) with longitudinal speed is justified. Consequently, SAMFC proposes the following adaptation law:

$$\alpha(t_k) = \begin{cases} \alpha_0 & \text{if } v_x(t_k) < v_{x,0} \\ K_\alpha \cdot (v_x(t_k) - v_{x,0}) + \alpha_0 & \text{if } v_x(t_k) \geq v_{x,0} \end{cases} \quad (17)$$

where $\alpha$ has a lower bound at $\alpha_0$, which is kept from zero velocity up to a given speed $v_{x,0}$, and then it is proportionally increased to longitudinal speed variation with a constant slope $K_\alpha$.

In this paper, the implemented SAMFC controller derives from a second order iPD controller, so the control action $u_{SA-iPD}$ is defined as:

$$u_{SA-iPD}(t_k) = \frac{-\hat{F}(t_k) + \ddot{y}_{1r}(t_k) + K_p e(t_k) + K_d \hat{\dot{e}}(t_k)}{\alpha(t_k)} \quad (18)$$

where $\alpha(t_k)$ is defined in (17) and the tunable parameters are $K_p, K_d, \alpha_0, v_{x,0}, K_\alpha$.

*4.4. PID*

PID control is a well-known framework that has been applied widespread in the industry due to its implementation ease and the intuitive relationship between its control parameters and their effect on the response. Although it is not frequently used in vehicle lateral control on its own, there exist some examples that use a PID with modifications, e.g.: a PID with a simple gain scheduler is applied in Zainal et al. (2017); a pure pursuit plus PI structure is applied in Park et al. (2014); a Fractional Order PID is applied in Dong et al. (2021); and an adaptive PID control structure is applied in Zhao et al. (2012).

In this paper, a PID in parallel form is implemented with a filtered derivative to reduce the measurement noise, where the control action $u_{PID}$ is expressed as a $\mathcal{Z}$-transform function:

$$U_{PID}(z) = \left( K_p + K_i \frac{T_s \cdot z^{-1}}{1 - z^{-1}} + K_d \frac{N_{PID}}{1 + N_{PID} \frac{T_s \cdot z^{-1}}{1 - z^{-1}}} \right) E(z) \quad (19)$$

being the gains $K_p$, $K_i$, $K_d$ and the filter parameter coefficient $N_{PID}$ the tunable parameters.

*4.5. MPC*

Model Predictive Control (MPC) has been successfully applied in lateral vehicle control, but in the literature there is no consensus on the best approach for this purpose (see Stano et al. (2022) and the references therein). The objective of this work is not developing a new MPC structure, but using a simple-to-implement MPC algorithm that is robust and takes non-linearities into consideration — similar to strategies such as Kebbati et al. (2021), Lin et al. (2019) and Mata et al. (2019).

The MPC structure used in this work is a Nonlinear MPC (NLMPC), as it consist on an adaptive MPC structure where the dynamic model of the vehicle is considered Linear Time Varying (LTV) with the longitudinal speed of the vehicle as a variable parameter. In this implementation, a linear model is obtained for each control interval from (7) by fixing the current longitudinal speed; this model is used across the prediction horizon.

As usual in the MPC framework, a prediction horizon of $h_p$ sample instants is considered to predict future system states and a control horizon of $h_c$ sample instants is considered to make the system reach steady state.

The NLMPC controller receives the four states $\{e_y, \dot{e}_y, e_\psi, \dot{e}_\psi\}$ as measured variables, being the derivatives of the lateral and angular errors obtained directly, although only the lateral error is considered as the output to be regulated. In simulation tests, it was found that directly measuring all the states of the system model (7) enhances the positive-definiteness of the Quadratic Programming (QP) Hessian matrix of the controller, regardless of the control and prediction horizons used, preventing the optimization problem to become unsolvable and thus enhancing the controllability.

Note that the NLMPC strategy has been implemented with the aid of MATLAB's Nonlinear MPC functions. In this implementation, the maximum number of iterations allowed by the QP solver *fmincon* is 10 in order to reduce computation time, which can cause a loss of tracking quality. To mitigate the effect of the iteration limit, optimal prediction and control sequences generated every control period are used as initial guesses in the next one.

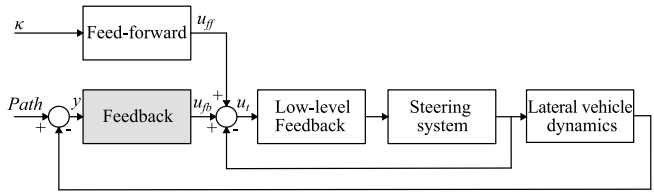

Fig. 1. Control scheme.

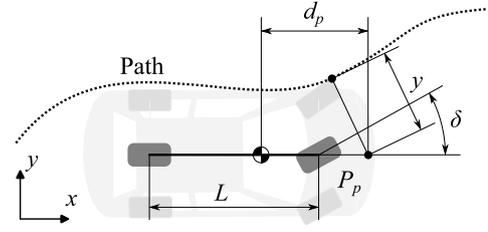

Fig. 2. Kinematic single-track model.

The cost function in the MPC optimization step is expressed with respect to the QP decision variable $z_k$:

$$z_k = \begin{bmatrix} u(k|k) & u(k+1|k) & \cdots & u(k+h_p-1|k) \end{bmatrix}^T \quad (20)$$

where $u$ is the control action (usually referred to as manipulated variable in the MPC framework) and $h_p$ is the prediction horizon. The function to be minimized $J(z_k)$ is the sum of a term related to the system output and another one associated to the time derivative of the control action:

$$J(z_k) = J_y(z_k) + J_{\dot{u}}(z_k) \quad (21)$$

The cost related to the system output is defined as the Root Mean Square Error (RMSE) between the system output $y$ and its reference $y_r$ along the prediction horizon $h_p$:

$$J_y(z_k) = \sum_{i=1}^{h_p} \left( y_r(k+i|k) - y(k+i|k) \right)^2 \quad (22)$$

The cost related to the derivative of the control action is defined as:

$$J_{\dot{u}}(z_k) = w_{\dot{u}} \sum_{i=0}^{h_c} (u(k+i|k) - u(k+i-1|k))^2 \quad (23)$$

where the weight of the manipulated variable rate $w_{\dot{u}}$ is varied to consider the smoothness of the control action and where $h_c$ is the control horizon.

Moreover, the control action and its rate are limited by the maximum steering angle and the maximum steering speed to prevent saturation of the steering actuator.

The NLMPC tunable parameters are $h_p$, $h_c$ and $w_{\dot{u}}$.

## 5. Evaluation methodology

In order to compare the different controllers as fairly as possible, a systematic procedure has been used and is the described below. First, a shared comparison benchmark is defined for all controllers, including the control scheme, the reference trajectories and the evaluation metrics. In addition, both the tuning procedure used for all controllers and the subsequent robustness analysis is described. Finally, the criteria applied to select the controller setups used in experimental tests is introduced.

### 5.1. Benchmark description

#### 5.1.1. Control scheme

The feedback control techniques introduced in Section 3 are compared under the control scheme shown in Fig. 1.

The low-level feedback controller uses a PD structure ($K_{llp} = 18$, $K_{lld} = 5$) to regulate the angular position of the steering wheel actuator and receives an angular position reference as input.

Table 1
Vehicle dynamic parameters.

| Parameter | $m$ [kg] | $I_z$ [kg m$^2$] | $C_f$ [N/rad] | $C_r$ [N/rad] | $l_f$ [m] | $l_r$ [m] |
|---|---|---|---|---|---|---|
| Value | 1372 | 1990 | 37 022.5 | 35 900 | 0.98 | 1.48 |

The outer part of the control scheme is composed of a feed-forward term and a feedback term. The feed-forward term introduces an anticipatory action considering the path curvature of the planned path. The feedback term deals with external disturbances or unmodeled dynamics in the feedforward model. This block is the focus of the comparative evaluation presented in this paper. As a result, the total steering wheel control action $\delta_t$ can be expressed as:

$$\delta_t = \delta_{max} \cdot \left( u_{ff} + u_{fb} \right) \quad (24)$$

where $u_{fb}$ is the feedback control action injected by each controller and $u_{ff}$ the feedforward component. The former was already applied in Godoy et al. (2015) and Moreno-Gonzalez et al. (2022) and relies on a kinematic model of the vehicle and the curvature of the path, as Fig. 2 shows, where $y_1$ is the lateral deviation at the preview point $P_p$ and $d_p$ is the preview distance. Note that speed-based variable preview distance is considered, so that $d_p = d_{p,0} + v_x \cdot t_p$, where $d_{p,0}$ is the minimum preview distance and $t_p$ is the preview time.

The feed-forward control term is defined as:

$$u_{ff} = \frac{R_S}{\delta_{max}} \cdot \arctan \left( L \cdot \kappa \right) \quad (25)$$

where the resulting control action is normalized ($u_{ff} \in [-1, 1]$), $L$ is the wheelbase, $\kappa$ is the path curvature, $R_S$ is the steering ratio and $\delta_{max}$ is the maximum steering angle.

Note that $d_{p,0}$ and $t_p$ are considered tunable parameters for all the implemented controllers.

With regard to vehicle parameters, Table 1 contains the values considered in the model-based controllers of this work, which have been estimated from the experimental platform used in the real tests.

#### 5.1.2. Benchmark trajectories

Six different trajectories have been used for both tuning and experimental evaluation. Each trajectory is targeted to a specific driving purpose: quite, moderate, aggressive-medium speed, high speed and aggressive-high speed. The driving purpose is not only reflected in the shape of the path but also in the speed and acceleration limit values used to compute the speed profiles, as shown in Table 2. For the computation of the speed profiles, the acceleration-limited speed planning algorithm presented in Artuñedo et al. (2022) has been used. The benchmark trajectories are shown in Fig. 3.

#### 5.1.3. Performance metrics

On the one hand, the integral absolute lateral error (IAE) is used to evaluate the tracking quality. On the other hand, as the classical IAU (Integral of the Absolute value of the control signal) is too simplistic to properly analyze the system dynamics, the frequency spectrum of the feedback control action is used to define two different performance indicators:

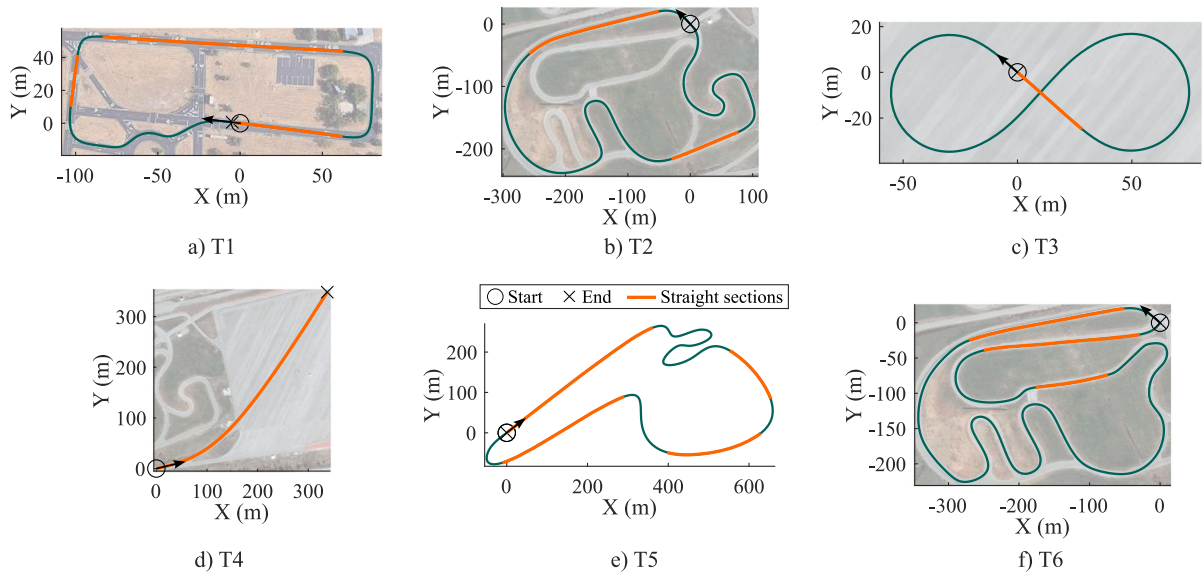

**Fig. 3.** Benchmark trajectories. (For interpretation of the references to color in this figure legend, the reader is referred to the web version of this article.)

**Table 2**
Trajectories used in optimization and experimental tests.

| Trajectory | T1 | T2 | T3 | T4 | T5 | T6 |
|---|---|---|---|---|---|---|
| Maximum speed (km/h) | 35 | 71 | 66 | 120 | 100 | 70 |
| Maximum longitudinal acceleration (m/s$^2$) | 0.4 | 1.0 | 2.2 | 2.5 | 1.5 | 2.0 |
| Maximum longitudinal deceleration (m/s$^2$) | 0.7 | 2.0 | 3.0 | 3.5 | 2.0 | 2.0 |
| Maximum lateral acceleration (m/s$^2$) | 1.0 | 2.0 | 4.0 | 4.0 | 4.0 | 2.0 |
| Length (m) | 471.0 | 1391.8 | 354.3 | 500.0 | 2119.6 | 1959.3 |
| Driving purpose | Quite | Moderate | Aggressive-medium speed | High speed | Aggressive-high speed | Moderate |
| Usage (O: Optimization, T: Testing) | O/T | T | T | T | O | O |

1. $M_\epsilon$: this metric quantifies the low frequency oscillations of the control action, which can lead to vehicle instability.
2. $M_\zeta$: this variable quantifies the high frequency oscillations of the control action, which do not destabilize the system per se, but cause great discomfort to the vehicle occupants.

The values of both metrics, $M_\epsilon$ and $M_\zeta$, are computed from two separated frequency bands, experimentally identified: $\epsilon$ (1.1–4 Hz) and $\zeta$ (4–10 Hz), respectively. Note that a control frequency of 20 Hz is assumed in this work. A high-pass filter is firstly applied to the feedback control action to remove undesirable spectral power values at low frequencies. The cutoff frequency of these filters are 0.5 Hz and 4 Hz for $M_\epsilon$ and $M_\zeta$, respectively. Then, the spectrum is calculated by applying the Short-time Fourier transform with 5-s overlapping sections.

The value of $M_\epsilon$ is finally calculated as the mean of the maximum power spectrum at each section, considering a threshold and a scale factor to balance the order of magnitude of both metrics:

$$M_\epsilon = \frac{1}{n}\sum_{i=1}^{n} s_\epsilon \cdot \max\left(10 * \log P_{\epsilon,i} + \lambda_\epsilon\right)$$

where $n$ is the amount of 5-s sections, $P_{\epsilon,i}$ is the spectrum power of band $\epsilon$ in section $i$, $s_\epsilon$ is a scale factor ($s_\epsilon = 0.015$) and $\lambda_\epsilon$ is a threshold in dB ($\lambda_\epsilon = 80$ dB).

The high sensitivity of $M_\epsilon$ may generate wrong metric values in curves, given its low-frequency spectrum. Hence, only straight and long sections where the path curvature is below $0.01 \, \text{m}^{-1}$ and the vehicle drives for more than 5 s (considering the speed profile of each trajectory) are considered. The straight sections for each test trajectory are drawn in orange in Fig. 3.

The procedure followed to obtain $M_\epsilon$ is also applied for $M_\zeta$. However, instead of the mean, the maximum power among all sections is assigned to this metric, using a scale factor of $s_\zeta = 0.04$ and the same threshold $\lambda_\zeta = 80$ dB. This particularity is motivated by the low equivalence found in experimental tests between what is intended to be reflected by this metric and the value obtained when the mean value is used. However, when the maximum power is considered, controllers that exhibit high frequency oscillations in any section of the test trajectory are penalized with high values of $M_\zeta$.

To summarize, IAE is the chosen indicator of the reference tracking quality, $M_\epsilon$ measures the (in)stability margin of the controller and $M_\zeta$ reflects passenger discomfort.

### 5.2. Tuning procedure

The tuning methodology used in a comparative evaluation of path-following controllers is of great importance. In order to make the comparison between control approaches as fair as possible, this paper proposes different metrics that are used in a multi-objective optimization, as described below.

Due to the existence of a wide parameter research space, in which there are infinite combinations of the different acceptable performance metrics, it is not easy to compare between different control structures. Indeed, some may have a faster response than others in detriment of the control action metrics, which may be better for the latter. Moreover, the required tracking accuracy and control action safety and softness vary depending on the vehicle speed and driving dynamics. Therefore, it is not straightforward to determine beforehand the importance of each metric and, consequently, it is not possible to define a universally valid cost function as a unique weighted sum of the metrics of tracking quality and control effort. Besides, inferring a set of control parameters from a precise value of a control action metric (namely $M_\epsilon$ or $M_\zeta$) is not straightforward. As a result, the following multi-objective optimization problem is posed:

$$\min \mathbf{J}(\mathbf{x},\mathbf{p})$$

s.t. $\dot{\mathbf{x}} = f(\mathbf{x}, \mathbf{p})$  (26)

$\mathbf{p} \in [\mathbf{p}_{min}, \mathbf{p}_{max}]$  (27)

where the set of objective functions $\mathbf{J}(\mathbf{x}, \mathbf{p})$ is defined as $(\max\{IAE_i\}, \max\{M_{\epsilon,i}\}, \max\{M_{\zeta,i}\})^T$, being $i$ the number of benchmark trajectory $T_i$; and $\mathbf{p}$ is the set of tunable parameters of the control structures that will be compared in Section 6. The functions to minimize are constrained by the system dynamics (26) and the control parameters are bounded by design by (27). The simulation model employed in the optimization process accurately replicates the experimental vehicle utilized in this work. For this purpose, a dynamic model with 14 degrees of freedom (6 for the vehicle body motion: longitudinal, lateral, vertical, roll, pitch, and yaw; and 2 for each wheel: vertical motion and spin) has been used. The power-train modeling comprises three elements: (i) the engine, whose torque map has been modeled from measurements taken in the experimental platform; (ii) the gearbox, which includes the same drive ratios and gear shifting logic than the real vehicle, and (iii) the resistance toques coming from braking system, longitudinal wind forces and gravitational forces. The tire behavior was reproduced with the Pacejka tire model (Pacejka & Bakker, 1992).

To solve the multi-objective optimization problem, a Pareto Efficiency test is carried out using MATLAB's ParetoSearch algorithm (Custódio et al., 2011), as it allows to search for combinations of control parameters with optimal responses considering different goals. By running this method, a Pareto front is obtained for each controller that shows its potential to minimize each of the objectives, i.e., the performance metrics. Moreover, the sets of control parameters that generate the Pareto front are known, so the relationship between control parameters and metrics can be studied. This algorithm automatically searches for optimal response points across the control parameter space, modifying the parameters up to a user-defined tolerance.

The work zone in the Pareto front has been defined after multiple tests, in which it is observed that: (i) an IAE greater than 0.35 m implies poor tracking at the curves, (ii) an $M_\epsilon$ greater than 0.25 can lead the system to instability when the dynamic constraints are varied, and (iii) an $M_\zeta$ greater than 0.7 leads to a loss of passenger comfort.

To make the tuning results more general, every controller is simulated in three of the six benchmark trajectories: T1, T5 and T6. These trajectories have been selected in order to cover a wide operation range. The three performance metrics are obtained for each trajectory, but only the maximum of each metric is considered in the optimization to ensure that the selected controller parameterization will be as much stable, comfortable and accurate as possible in the most unfavorable situation.

The result of the tuning procedure are depicted in Fig. 4. This figure shows 3 different views of the Pareto front obtained for each controller. In this figure each point represents a parameter setup. More details on the numbering and notation used in these charts is provided at the end of Section 5.4.

### 5.3. Robustness tests in simulation

The optimization described above is used to generate a Pareto front with regard to IAE, $M_\epsilon$ and $M_\zeta$. However, the Pareto diagram does not provide explicit information on whether the control configurations that are part of the front bring the vehicle close to instability when disturbances and/or critical parameters (e.g., mass, friction) variations arise.

Given the impossibility of performing numerous tests in real environments where difficult-to-estimate parameters, such as those related to tire and road physics, are varied, the robustness tests have been carried out in simulation using the Monte Carlo method. In order to consider driving over a wide enough range of disturbance variation to cover extreme but realistic scenarios, in this analysis four parameters are randomly varied for each simulation: (i) the vehicle mass using a normal distribution with $\mu_m = 1372\,\mathrm{kg}$ and $\sigma_m = 137.2\,\mathrm{kg}$; (ii) the vehicle inertia around the Z axis ($I_z$) using a normal distribution with $\mu_{I_z} = 1990\,\mathrm{kg\,m^2}$ and $\sigma_{I_z} = 199.0\,\mathrm{kg\,m^2}$; (iii) the static friction coefficient using an uniform distribution between 0.5 and 1.17 and (iv) the lateral stiffness-slip factor (typically named as "a3" parameter of Pacejka tire model Pacejka & Bakker, 1992) using a normal distribution with $\mu_{a3} = 80157\,\mathrm{N/rad}$ and $\sigma_{a3} = 16031\,\mathrm{N/rad}$. Note that this distributions are selected in order to represent the usual variations found in these parameters considering their probability of occurrence in realistic scenarios.

To that end, 200 random parameter variations have been performed for each controller setup belonging to the Pareto front of each control structure. Thus, a total of 217 400 simulations were run using trajectory T5 since it is the most demanding trajectory in terms of speed and longitudinal and lateral acceleration.

The criterion to classify a simulation as valid is to reach the end of the trajectory without exceeding a lateral error of 3 m during the entire trajectory. This limit has been stated from experimental tests to automatically detect unacceptable lateral error and instability during the simulation.

The results of this evaluation are shown in Fig. 5, where a color scale is used to represent the percentage of success of each controller setup, i.e., the amount of valid simulations over the total simulations for each controller setup. As can be seen, in general, the controller parameters sets with lower levels of stability have high values of some of the 3 objective metrics used in the Pareto front, as expected. For example, for MFC controller, it can be seen in Fig. 5(b) that setups that reach lower values of IAE, i.e., the best tracking quality, present a low success percentage caused by a greater control action aggressiveness to increase the tracking quality that is leading the vehicle to instability in some of the simulations.

### 5.4. Controller setups selection criteria

Since the shape of the Pareto front differs according to the control technique, a selection rule using fixed values of the metrics cannot be established. Thus, to choose comparable parameterizations among controllers, the following criteria have been defined:

First of all, the selection area is constrained to maximum assumable values for each metric/axis that have been determined from experimental tests (see Section 5.2). Moreover, a minimum rate of 90% of success in the robustness tests carried out in Section 5.3 is set. Among the Pareto front region considering these limits, 3 different sets of parameters have been selected for each controller: Setup 1 is the one with minimum value of $M_\epsilon$ among the 5 points with minimum IAE. Setup 3 is the one that minimizes $M_\epsilon$ among the 5 points with maximum IAE. Finally, Setup 2 is selected as the closest point to the bisector of the angle formed between the vectors joining the origin of coordinates with Setup 1 and Setup 3.

The resulting setups selected for real tests once the aforementioned criteria is applied are highlighted in Fig. 4 using a circle filled with the same color used for each controller. The setup number is specified next to each circle. The parameters belonging to each setup for each controller are specified in Table 3.

## 6. Experimental results and discussion

### 6.1. Experimental platform

The comparative evaluation was carried out using one of the automated vehicles of the AUTOPIA group (see Fig. 6) at the test track of the Centre for Automation and Robotics (CSIC-UPM) in Arganda del Rey, Spain.

The localization of the vehicle used in this work relies on a RTK-GNSS receiver and on-board sensors to measure vehicle speed, accelerations and yaw rate. The vehicle also includes a computer with an

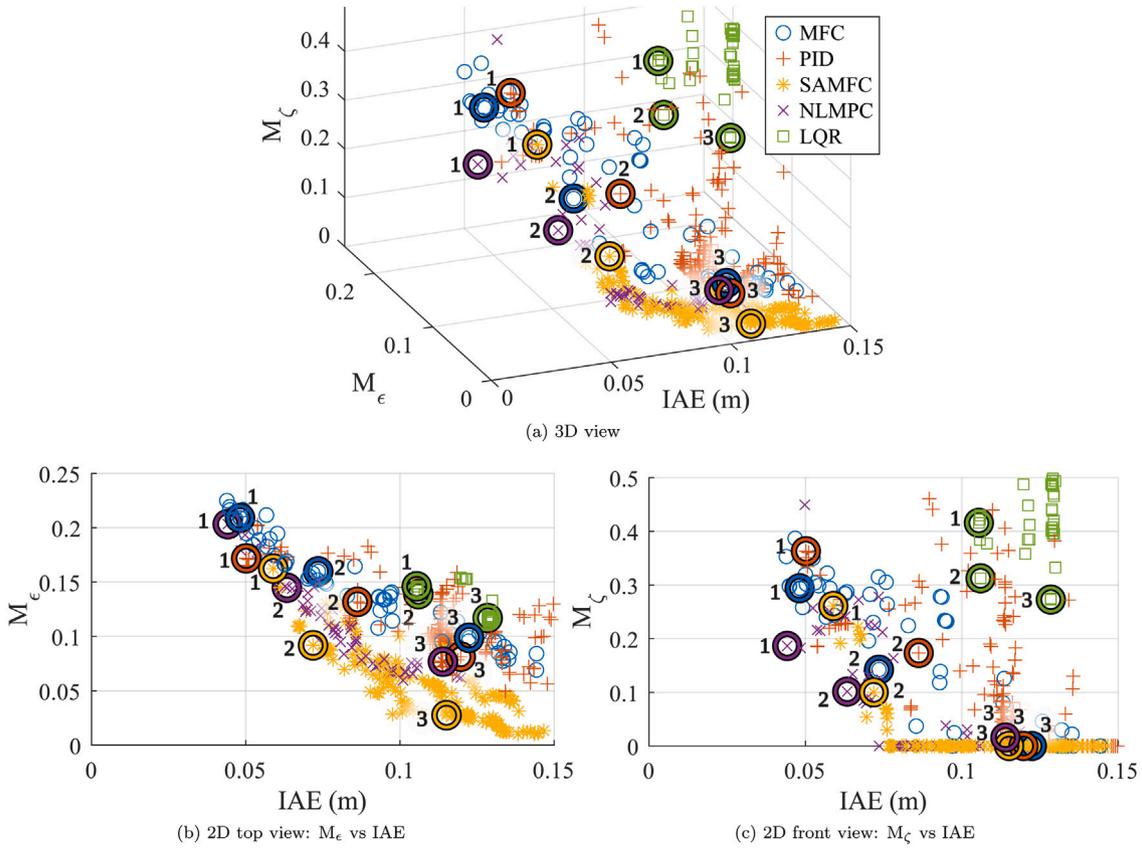

(a) 3D view

(b) 2D top view: $M_\epsilon$ vs IAE

(c) 2D front view: $M_\zeta$ vs IAE

**Fig. 4.** 3D Pareto front: multi-objective optimization results for each controller evaluated. (For interpretation of the references to color in this figure legend, the reader is referred to the web version of this article.)

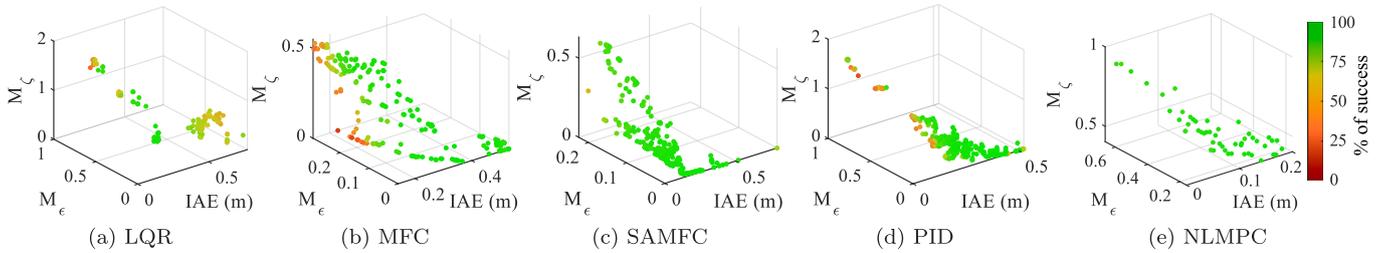

(a) LQR  (b) MFC  (c) SAMFC  (d) PID  (e) NLMPC

**Fig. 5.** Monte Carlo test results. (For interpretation of the references to color in this figure legend, the reader is referred to the web version of this article.)

Intel Core i7-8700T and 32 GB RAM, which is used to run the feedback control algorithms (Artuñedo et al., 2019) compared in this work. This software architecture runs on a soft real-time Linux-based operating system.

The trajectory tracking system used in this work is designed and behaves in a decoupled manner: on the one hand, a longitudinal controller computes the positions of throttle and brake pedals from the reference speed profile and the speed error. On the other hand, the lateral controller uses the lateral error measured from the ego-vehicle pose with respect to the reference path, to compute the steering wheel position. Both lateral and longitudinal control actions are computed at a frequency of 20 Hz. Dedicated digital positioning controllers are in charge of low-level control tasks.

### 6.2. Experimental results

In order to evaluate the controllers in real driving environments, extensive experimental tests have been carried out in four different trajectories: T1, T2, T3 and T4 (see Fig. 3), thus covering driving styles ranging from quite to aggressive and high-speed. This subsection provides an in-depth analysis of the results in terms of tracking quality, the metrics $M_\epsilon$ and $M_\zeta$, and computation time. The experimental tests comprise the evaluation of the 15 controller setups shown in Table 3 in trajectories T1, T2 and T3. Moreover, among the three setups for each control structure, the one with minimum IAE in the Pareto front is evaluated in T4. Hereinafter, these configurations will be referred to as "the best setup of each controller".

Table 4 shows the numeric results of the integral absolute value (IAE) of the lateral error, the maximum lateral error (MLE) and the metrics $M_\epsilon$ and $M_\zeta$. In addition to the results for each setup, a row including the mean values of the 3 setups for each control technique and a column with the mean values of each setup in all trajectories are added. To better represent the values, a color scale from green to red has been used to represent small and large values of each metric respectively. Moreover, the minimum values for each metric in each trajectory is highlighted in bold as well as the minimum mean values in the gray rows.

The first point to highlight from the results is that the metric values of all setups remains below the assumable thresholds stated in Section 5.2. Overall, the NLMPC and SAMFC are the controllers that perform better across multiple performance metrics and trajectories. On the one hand, NLMPC consistently achieves the lowest values in terms of IAE. NLMFC-1 setup yields the lowest IAE in all trajectories. On the other hand, SAMFC keeps low values of IAE while obtaining lower values of MLE in T1 (SAMFC-2) and $M_\epsilon$ in T2 (SAMFC-2).

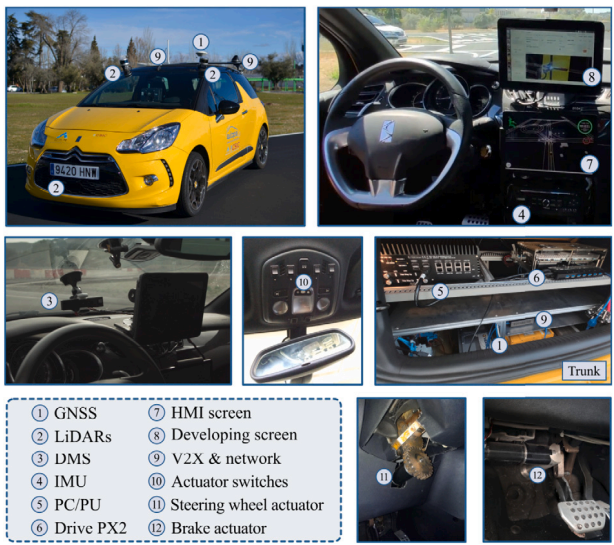

**Fig. 6.** Experimental platform.

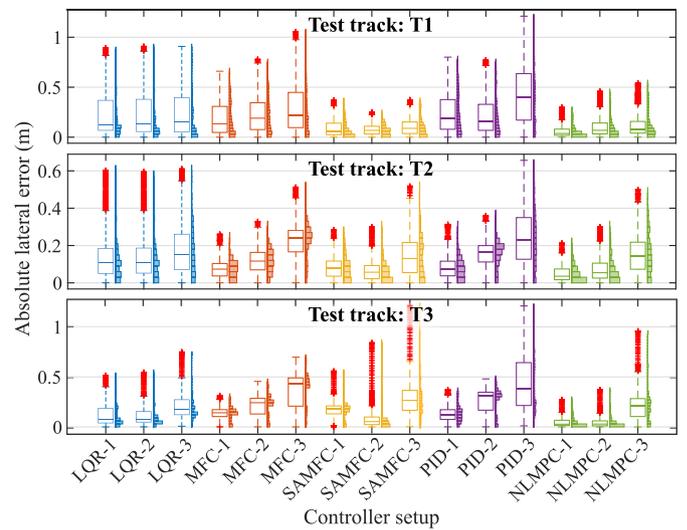

**Fig. 7.** Box plots of the lateral error during tests on T1, T2 and T3.

Table 3
Parameters selected from Pareto front.

| Controller | Parameters | Setup | | |
|---|---|---|---|---|
| | | 1 | 2 | 3 |
| LQR | $q_1$ | 0.002 | 0.002 | 0.001 |
| | $q_2$ | 0.0002 | 0.0002 | 0.0002 |
| | $q_3$ | 0.001 | 0.001 | 0.001 |
| | $q_4$ | 0.0002 | 0.0002 | 0.0001 |
| | $d_p$ | 0.000 | 0.000 | 0.000 |
| | $N$ | 6.158 | 9.543 | 7.911 |
| MFC | $k_p$ | 0.000 | 0.000 | 0.000 |
| | $k_d$ | 3.337 | 3.603 | 1.810 |
| | $\alpha$ | 373.2 | 502.4 | 373.8 |
| | $d_p$ | 1.516 | 1.149 | 0.516 |
| SAMFC | $k_p$ | 0.000 | 0.750 | 0.125 |
| | $k_d$ | 4.266 | 2.766 | 2.141 |
| | $\alpha_0$ | 94.4 | 93.6 | 93.0 |
| | $d_p$ | 1.000 | 0.625 | 0.000 |
| | $v_{x,0}$ | 2.68 | 12.78 | 10.66 |
| | $K_\alpha$ | 10.0 | 10.0 | 10.0 |
| PID | $k_p$ | 0.160 | 0.153 | 0.071 |
| | $k_d$ | 0.030 | 0.065 | 0.027 |
| | $k_i$ | 0.000 | 0.000 | 0.000 |
| | $d_p$ | 1.763 | 0.059 | 2.346 |
| | $N$ | 8 | 20 | 3 |
| NLMPC | $h_p$ | 11 | 13 | 21 |
| | $h_c$ | 3 | 4 | 3 |
| | $w_{\ddot{u}}$ | 15.00 | 26.08 | 43.11 |
| | $d_p$ | 0.000 | 0.234 | 0.078 |

It is worth noting that for the MLE metric, the LQR and PID have significantly higher values compared to the other setups, specifically the third setup of each controller. Moreover, both PID and LQR deliver a poor tracking quality in comparison with NLMPC, SAMFC and MFC. Similarly, for the $M_\epsilon$ metric, the LQR-2 setup has a lower value compared to the others. Nevertheless, the resulting values of $M_\epsilon$ for all tested controllers and setups are assumable.

Looking at the mean column, we can see that the NLMPC setup has the lowest average values across all the metrics, including $M_\zeta$. The values of $M_\zeta$ are quite low for all controllers except for PID-2, which presents a value close to the upper limit considered as assumable (0.676 over 0.7). The MFC setup has relatively higher mean values IAE, MLE and $M_\zeta$. This aligns with the individual performance analysis we conducted earlier.

Comparing LQR and MFC, it is observed that the MFC controller setup performs relatively well in trajectories T1 and T2 in terms of tracking quality, both keeping low values of $M_\epsilon$. However, at T3, the LQR achieves a lower IAE. Note also that the MFC controller shows better performance in terms of MLE and $M_\zeta$.

With regard to tracking quality, IAE provides useful information on the average tracking performance during the entire test. However, to further analyze the occurrence of the different lateral error magnitudes, Fig. 7 shows box plots of the lateral error for each controller configuration in Table 4. Note that a histogram is shown on the right of each box plot for a better representation of the lateral error distribution for each controller setup. It can be observed that, depending on the context, some controllers perform better in terms of tracking quality: In T1, the quietest of the trajectories, it is observed how all controllers are able to concentrate most of the error measurements close to 0. However, when the trajectories are more demanding (T2 and T3), differences in the error distribution are observed: NLMPC and SAMFC are still able to keep most of the error measurements close to 0 while the PID lateral error distribution worsens considerably.

In view of the results in Pareto front (Fig. 4), SAMFC-2 was expected to have a greater IAE than SAMFC-1 and lower than SAMFC-3. Nevertheless, from the experimental results we can see that the lowest IAE of the SAMFC setups is found in setup 2. Nevertheless, Fig. 7 shows that although the IAE is smaller in SAMFC-2 than in SAMFC-1 in all trajectories, the lateral error is distributed over a wider range, i.e., the MLE is larger in SAMFC-1 in trajectories T2 and T3. For the rest of the controllers, the increasing trend of the IAE from setup 1 to 3 observed in the Pareto front, is also noted in the real results.

Fig. 8 shows the box plots and histograms of the lateral error of the setup with lower IAE of each control technique during the tests on all testing trajectories (T1–T4). This figure shows the differences in the error distribution between the various trajectories. In trajectory T1, devoted to quiet driving, it can be seen that SAMFC-2 and NLMPC-1 obtain a similar IAE. However, SAMFC-2 manages to concentrate the error in a narrower range. Although the rest of the controllers behave well, they obtain a higher lateral error. At T2, focused on moderate driving, LQR-1 clearly performs worse than the rest. The T3 trajectory is the most aggressive in terms of maximum lateral and longitudinal accelerations. Among the metrics, the most significant disparity when compared to T1, T2, and T4, emerges specifically in the maximum lateral error achieved by SAMFC-2. This controller exhibits a notably higher value in this regard compared to the other controllers. However, it is able to achieve a very low IAE, which is only surpassed by NLMPC-1. Finally, at T4, focused on high speed, it is noted that LQR-1 achieves higher errors than the rest, with MFC-1, SAMFC-2 and PID-1 being similar and NLMPC-1 being slightly lower.

**Table 4**
Results on T1, T2 and T3 of 3 setups for each controller.

| Controller | T1 | | | | T2 | | | | T3 | | | | Mean (T1, T2, T3) | | | |
|---|---|---|---|---|---|---|---|---|---|---|---|---|---|---|---|---|
| setup | $IAE$ | $MLE$ | $M_\epsilon$ | $M_\zeta$ | $IAE$ | $MLE$ | $M_\epsilon$ | $M_\zeta$ | $IAE$ | $MLE$ | $M_\epsilon$ | $M_\zeta$ | $IAE$ | $MLE$ | $M_\epsilon$ | $M_\zeta$ |
| LQR-1 | 0.239 | 0.893 | 0.010 | 0.322 | 0.146 | 0.603 | 0.076 | 0.286 | 0.128 | 0.516 | 0.231 | 0.152 | 0.171 | 0.671 | 0.106 | 0.253 |
| LQR-2 | 0.239 | 0.909 | 0.009 | 0.290 | 0.146 | 0.594 | 0.070 | 0.278 | 0.126 | 0.548 | 0.127 | **0.000** | 0.170 | 0.684 | 0.069 | 0.189 |
| LQR-3 | 0.249 | 0.909 | 0.005 | 0.169 | 0.178 | 0.611 | 0.063 | 0.293 | 0.224 | 0.750 | 0.093 | 0.061 | 0.217 | 0.757 | 0.054 | 0.174 |
| LQR[a] | 0.242 | 0.904 | **0.008** | 0.260 | 0.157 | 0.603 | **0.070** | 0.286 | 0.159 | 0.605 | 0.150 | **0.071** | 0.186 | 0.704 | **0.076** | 0.206 |
| MFC-1 | 0.184 | 0.660 | 0.070 | **0.000** | 0.074 | 0.261 | 0.157 | 0.296 | 0.138 | 0.315 | 0.151 | 0.108 | 0.132 | 0.412 | 0.126 | 0.135 |
| MFC-2 | 0.229 | 0.779 | 0.016 | 0.102 | 0.118 | 0.329 | 0.101 | 0.374 | 0.220 | 0.456 | 0.121 | 0.089 | 0.189 | 0.521 | 0.079 | 0.188 |
| MFC-3 | 0.294 | 1.057 | 0.026 | 0.135 | 0.223 | 0.513 | 0.090 | 0.297 | 0.380 | 0.697 | 0.145 | 0.571 | 0.299 | 0.756 | 0.087 | 0.334 |
| MFC[a] | 0.236 | 0.832 | 0.037 | 0.079 | 0.138 | 0.368 | 0.116 | 0.322 | 0.246 | **0.489** | **0.139** | 0.256 | 0.207 | 0.563 | 0.097 | 0.219 |
| SAMFC-1 | 0.093 | 0.372 | 0.077 | 0.355 | 0.085 | 0.284 | 0.100 | 0.232 | 0.175 | 0.560 | 0.356 | 0.214 | 0.118 | 0.405 | 0.178 | 0.267 |
| SAMFC-2 | 0.077 | **0.251** | 0.108 | 0.569 | 0.066 | 0.301 | 0.088 | 0.380 | 0.118 | 0.844 | 0.276 | 0.407 | 0.087 | 0.465 | 0.157 | 0.452 |
| SAMFC-3 | 0.112 | 0.375 | 0.048 | 0.287 | 0.144 | 0.520 | 0.061 | 0.522 | 0.315 | 1.209 | 0.334 | 0.365 | 0.190 | 0.701 | 0.148 | 0.391 |
| SAMFC[a] | **0.094** | 0.333 | 0.078 | 0.404 | **0.098** | 0.368 | 0.083 | 0.378 | 0.203 | 0.871 | 0.322 | 0.329 | 0.132 | 0.524 | 0.161 | 0.370 |
| PID-1 | 0.235 | 0.801 | 0.003 | 0.084 | 0.083 | 0.314 | 0.112 | 0.451 | 0.132 | 0.373 | 0.250 | 0.239 | 0.150 | 0.496 | 0.122 | 0.258 |
| PID-2 | 0.227 | 0.773 | 0.038 | 0.434 | 0.153 | 0.361 | 0.138 | 0.849 | 0.262 | 0.480 | 0.144 | 0.745 | 0.214 | 0.538 | 0.107 | 0.676 |
| PID-3 | 0.433 | 1.211 | **0.000** | **0.000** | 0.245 | 0.657 | **0.012** | 0.013 | 0.445 | 1.209 | 0.083 | **0.000** | 0.374 | 1.026 | **0.032** | 0.004 |
| PID[a] | 0.298 | 0.928 | 0.014 | 0.173 | 0.160 | 0.444 | 0.087 | 0.438 | 0.280 | 0.687 | 0.159 | 0.328 | 0.246 | 0.687 | 0.087 | 0.313 |
| NLMPC-1 | **0.063** | 0.300 | 0.082 | 0.134 | **0.050** | **0.214** | 0.165 | **0.000** | 0.051 | 0.274 | 0.419 | 0.569 | **0.055** | **0.263** | 0.222 | 0.234 |
| NLMPC-2 | 0.106 | 0.459 | 0.075 | **0.000** | 0.075 | 0.301 | 0.135 | **0.000** | 0.056 | 0.374 | **0.075** | **0.000** | 0.079 | 0.378 | 0.095 | **0.000** |
| NLMPC-3 | 0.127 | 0.542 | **0.000** | **0.000** | 0.157 | 0.501 | 0.052 | **0.000** | 0.242 | 0.959 | 0.145 | **0.000** | 0.175 | 0.667 | 0.066 | **0.000** |
| NLMPC[a] | 0.099 | 0.434 | 0.052 | **0.045** | 0.094 | **0.339** | 0.117 | **0.000** | **0.116** | 0.536 | 0.213 | 0.190 | **0.103** | 0.436 | 0.128 | **0.078** |

[a] Mean values of the 3 setups for each controller.

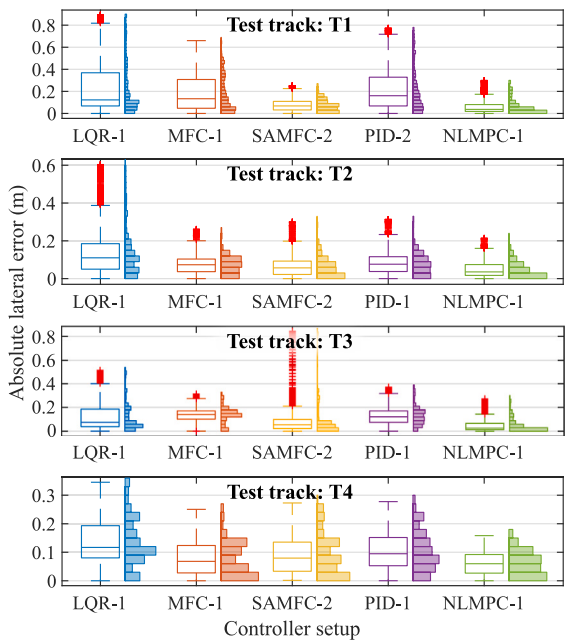

**Fig. 8.** Box plots of the best setup of each controller on T1, T2, T3 and T4.

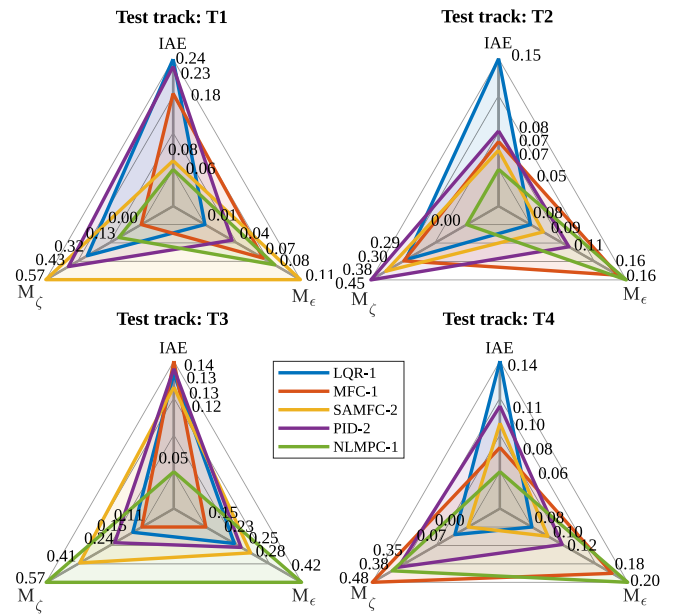

**Fig. 9.** Spider plots of IAE, $M_\epsilon$ and $M_\zeta$ of the best setup of each controller on T1, T2, T3 and T4.

In order to jointly represent the main metrics that are evaluated in this comparison, Fig. 9 uses spider graphs to show IAE, $M_\epsilon$ and $M_\zeta$ of the best setup of each controller on trajectories T1, T2, T3 and T4. It can be observed in T3 and T4 that a reduction of the IAE leads to an increase of the values of $M_\epsilon$ and $M_\zeta$, i.e., parameter sets that result in low tracking error, show a less comfortable driving and closer to instability. Although this behavior can be generalized, looking at T1 results in Fig. 9, it can be noted that NLMPC-1 is able to deliver a low value of $M_\zeta$ while achieving the lowest IAE. However, the $M_\epsilon$ value ranks as the second highest, behind that of SAMFC-2, which is also able to achieve a low IAE in T1. In the context of T2, NLMPC-1 once more secures the lowest IAE value, while also maintaining the smallest $M_\zeta$ value. However, it is worth noting that among all the controllers, NLMPC-1 achieves the highest $M_\epsilon$ value in this scenario.

To analyze in more detail the influence of the shape of the path on the occurrence of the different lateral error values, Fig. 10 shows box plots of the lateral error as a function of the curvature of the path. Note the vertical black dashed line indicating the value of $\kappa = 0\,\text{m}^{-1}$, i.e., the value corresponding to straight sections. In T1, SAMFC-2 and NLMPC-1 achieve very low lateral errors, but the errors increase in the curved areas, i.e., where the curvature is greater. Interestingly enough, the opposite is true for PID-2. Besides, in T3, it is observed that all the controllers present the highest lateral errors at the same curvature values. Note that the section of the trajectory where this occurs is the entrance to the second curve of the trajectory, where the vehicle arrives at high speed. Fig. 10 shows how MFC-1 and NLMPC-1 are able to handle the error better than LQR-1 and SAMFC-2 in these cases. This behavior can be further observed from the temporal plots of lateral error described below.

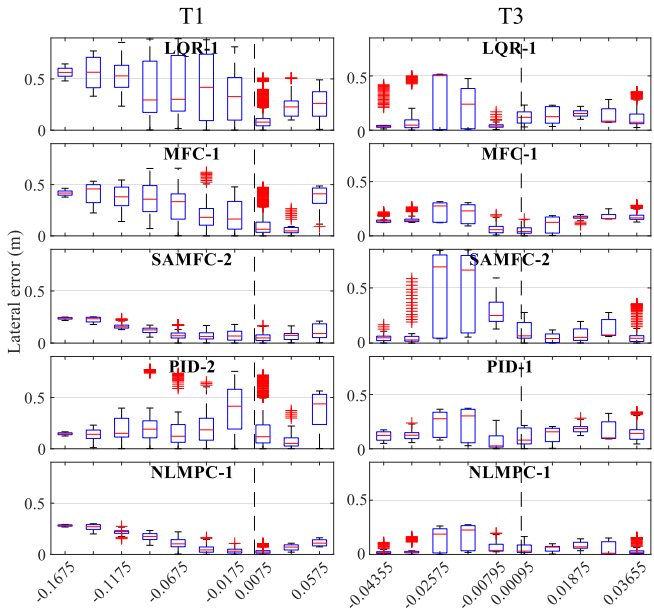

**Fig. 10.** Lateral error vs. curvature of the best setup of each controller on T1 and T3.

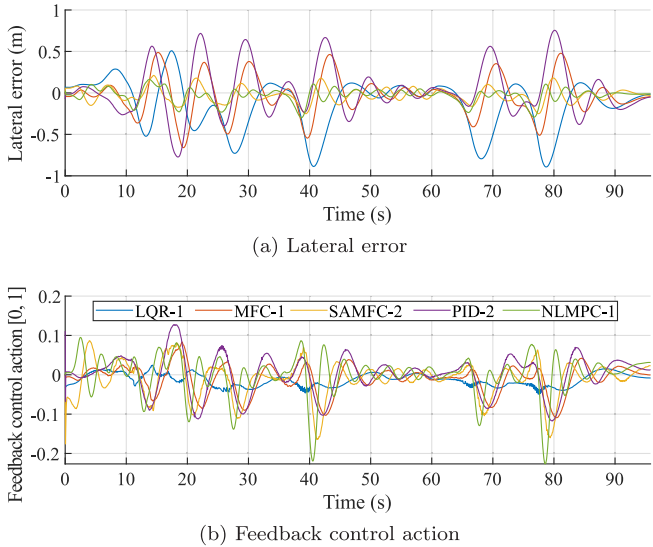

**Fig. 11.** Lateral error and feedback control action in T1.

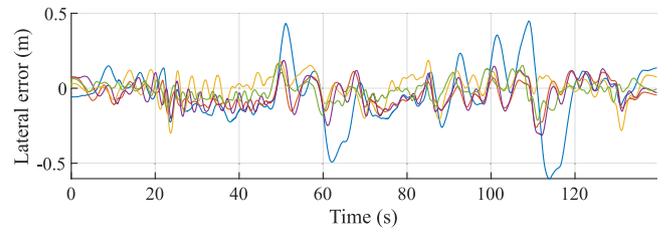
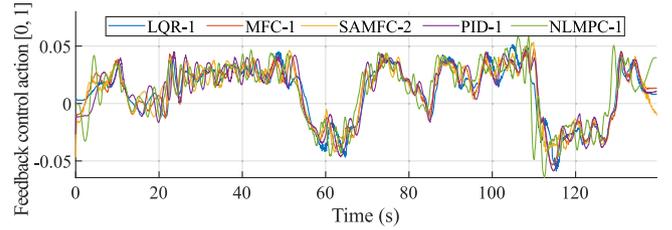

**Fig. 12.** Lateral error and feedback control action in T2.

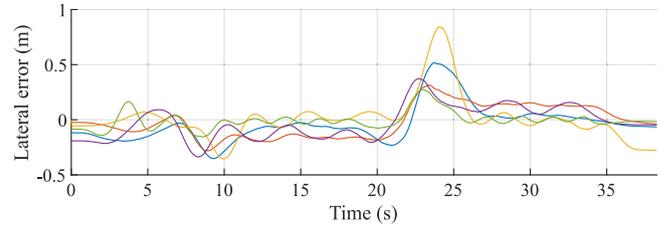
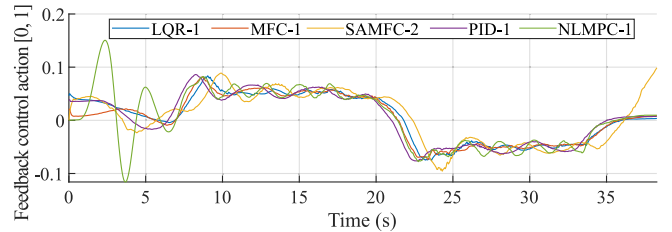

**Fig. 13.** Lateral error and feedback control action in T3.

The temporal evolution of lateral error and control action of the best controller configuration in trajectories T1, T2 and T3 are shown in Figs. 11, 12 and 13, respectively. In T1, the greatest absolute errors are reached by LQR and PID in presence of curves (e.g., instants $t = 40, 70$ and $80$ s). Nonetheless, it can be noted in Fig. 11(a) that LQR anticipates the curve causing the lateral error to occur on the inside of the curve. In contrast, the behavior of the PID is the opposite, reaching an error of similar magnitude but with the opposite sign, i.e., on the outside of the curve. This effect is not observed in the rest of the trajectories.

In T2 it can be seen that while all the controllers have good tracking until instant $t = 47$ s, the LQR-1 experiences significantly greater error values at instant $t = 50$ s than the other controllers. This instant corresponds to driving in a curve when the vehicle is traveling at $71$ km/h. Further on, it is also observed how the highest errors are obtained by LQR-1 in the rest of the curves of the trajectory, while the rest of the controllers are able to overcome this circuit, focused on moderate driving, without obtaining such large tracking errors.

With regard to T3, the most remarkable aspect in the temporal evolution of lateral error (Fig. 13(a)) is found at $t = 23$ s. At this instant, the controllers are facing the entrance to a sharp curve at the maximum trajectory speed: $66$ km/h. It is noted that although SAMFC tracking was excellent, it reaches at this point the maximum value of lateral error, followed by LQR. Note that in T3, in contrast to the results in T1 and T2, which are quieter trajectories, LQR-1 performs better showcasing the anticipatory behavior of LQR in demanding trajectories with high accelerations and speed.

In Fig. 14 the box plots of the control cycles runtime of all controllers during the tests performed in T1 are shown. Note that the most computationally expensive controller technique is NLMPC in comparison with LQR, PID, MFC and SAMFC. With regard to MPC, the runtime grows with the prediction horizon: a mean of 10.8 ms and standard deviation of 2.33 ms is obtained for a prediction horizon of 11 steps (NLMPC-1), while a mean of 25.3 ms and standard deviation of 8.27 ms is obtained for a prediction horizon of 21 steps (NLMPC-3). Note also that all other compared techniques are not affected by their parameters and keeps the runtime below 2 ms.

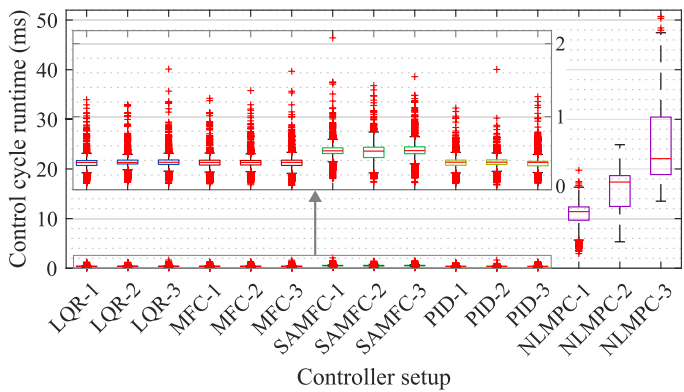

**Fig. 14.** Box plot of the runtimes of all controllers in T1 tests.

*6.3. Qualitative remarks*

Model-free approaches offer several benefits compared to the other controllers. Unlike controllers like LQR or MPC, which rely on system models, MFC and SAMFC operate without explicit knowledge of the underlying system dynamics. This model independence allows them to be more versatile and applicable to a wider range of systems that may include complex or uncertain dynamics, often hard to explicitly identify. Indeed, PID, MFC and SAMFC, being model-free approaches, typically require fewer assumptions and less system information compared to model-based controllers. This can simplify the design and implementation process and reduce the computational burden associated to some model-based control techniques, such as MPC. Nevertheless, the performance differences between PID, MFC and SAMFC are significant. Thus, SAMFC specifically highlights its speed-adaptive nature, suggesting that it has the ability to adjust its control parameters based on real-time system behavior. This adaptability allows SAMFC to deliver a good control performance even in dynamically changing driving environments. In view of the comparative results, SAMFC demonstrate robustness by consistently achieving competitive performance across multiple performance metrics. While it may not always have the absolute best performance in every metric, it performs well overall.

Model-based strategies such as MPC leverage the knowledge of the system dynamics and utilize mathematical models to design control strategies. This feature allows them to exploit the inherent understanding of the system behavior, leading to potentially better performance in scenarios where the system model is accurately known. However, the evaluation results show significant differences between LQR and MPC, being the latter able to better anticipate driving varying conditions.

The comparative results provide valuable insights into the selection of appropriate controllers for different driving purposes. Controllers like NLMPC-1 demonstrated versatility and robustness across various driving styles, while controllers like SAMFC-2 excelled in moderate driving scenarios.

The results of the tuning methodology applied revealed trade-offs between different performance metrics within each controller setup. For example, some controllers excelled in minimizing integral errors (IAE) but had higher $M_\epsilon$ or $M_\zeta$ values. This highlight the fact that optimizing one aspect of control performance may come at the expense of another.

Finally, it is worth noting the clear qualitative difference in terms of runtime between the NLMPC controller and the rest, according to Fig. 14. While the model employed for NLMPC in this comparison is relatively straightforward, the computational demands significantly exceed those of alternative techniques. Should a more intricate model be employed, the computational time constraints associated with NLMPC could pose challenges when deploying it in actual vehicles, where computational resources are often constrained.

## 7. Concluding remarks

This work provides a thorough examination of the strengths and weaknesses of various control formulations for the lateral control of an autonomous vehicle. By conducting an objective and comprehensive comparative analysis, this study offers valuable insights into the performance of relevant state-of-the-art control strategies.

The tuning procedure when comparing control strategies is a key aspect for a fair comparison, since the performance of a controller can radically change depending on its parameterization. In this work, a common tuning methodology has been followed for all the evaluated controllers. Two novel metrics are introduced to compare the controller behavior in terms of stability and comfort, providing a more comprehensive evaluation framework.

This comparison addressed both model-based and model-free approaches, ensuring a fair and diverse representation of control strategies. To validate the findings, extensive simulations and real tests were performed using an experimental instrumented vehicle, providing reliable and robust experimental data.

The comparative evaluation presented in this paper aims at serving as a valuable resource for researchers, engineers, and practitioners working in the field of autonomous driving. It assists in informed decision-making when selecting an appropriate control strategy for specific driving scenarios, taking into account various performance metrics and real-world driving scenarios.

**Declaration of competing interest**

The authors declare that they have no known competing financial interests or personal relationships that could have appeared to influence the work reported in this paper.

**Data availability**

Data will be made available on request.

**Acknowledgment**

We would like to express our sincere gratitude to the Spanish National Institute of Aerospace Technology (INTA) for their support allowing us to perform the real tests in their facilities.

**Appendix A. Supplementary data**

Supplementary material related to this article can be found online at https://doi.org/10.1016/j.arcontrol.2023.100910.

# References

Abdallaoui, S., Kribèche, A., & Aglzim, E. H. (2023). Comparative study of MPC and PID controllers in autonomous vehicle application. *Mechanisms and Machine Science*, *121*, 133–144.

Arifin, B., Suprapto, B. Y., Prasetyowati, S. A. D., & Nawawi, Z. (2019). The lateral control of autonomous vehicles: A review. In *ICECOS 2019 - 3rd international conference on electrical engineering and computer science, proceeding* (pp. 277–282). Institute of Electrical and Electronics Engineers Inc..

Artuñedo, A., Godoy, J., & Villagra, J. (2019). A decision-making architecture for automated driving without detailed prior maps. In *2019 IEEE intelligent vehicles symposium (IV)* (pp. 1645–1652).

Artuñedo, A., Villagra, J., & Godoy, J. (2022). Jerk-limited time-optimal speed planning for arbitrary paths. *IEEE Transactions on Intelligent Transportation Systems*, *23*(7), 8194–8208.

Balaji, T. S., & Srinivasan, S. (2023). Comparative study of different controllers in an autonomous vehicle system. *Materials Today: Proceedings*, *80*, 2390–2393.

Biswas, A., Reon, M. A. O., Das, P., Tasneem, Z., Muyeen, S. M., Das, S. K., Badal, F. R., Sarker, S. K., Hassan, M. M., Abhi, S. H., Islam, M. R., Ali, M. F., Ahamed, M. H., & Islam, M. M. (2022). State-of-the-art review on recent advancements on lateral control of autonomous vehicles. *IEEE Access*, *10*, 114759–114786.

Boyali, A., Mita, S., & John, V. (2018). A tutorial on autonomous vehicle steering controller design, simulation and implementation.

Calzolari, D., Schurmann, B., & Althoff, M. (2017). Comparison of trajectory tracking controllers for autonomous vehicles. In *2017 IEEE 20th international conference on intelligent transportation systems (ITSC)* (pp. 1–8). IEEE.

Chaib, S., Netto, M., & Mammar, S. (2004). H/sub /spl infin//, adaptive, PID and fuzzy control: a comparison of controllers for vehicle lane keeping. In *IEEE intelligent vehicles symposium, 2004* (pp. 139–144).

Chen, G., Yao, J., Hu, H., Gao, Z., He, L., & Zheng, X. (2022). Design and experimental evaluation of an efficient MPC-based lateral motion controller considering path preview for autonomous vehicles. *Control Engineering Practice*, *123*, Article 105164.

Custódio, A., Madeira, J., Vaz, I., & Vicente, L. (2011). Direct multisearch for multiobjective optimization. *SIAM Journal on Optimization*, *21*, 1109–1140.

Dominguez, S., Ali, A., Garcia, G., & Martinet, P. (2016). Comparison of lateral controllers for autonomous vehicle: Experimental results. In *2016 IEEE 19th international conference on intelligent transportation systems (ITSC)* (pp. 1418–1423).

Dong, X., Pei, H., & Gan, M. (2021). Autonomous vehicle lateral control based on fractional-order PID. In *2021 IEEE 5th information technology,networking,electronic and automation control conference (ITNEC), Vol. 5* (pp. 830–835).

Fliess, M., & Join, C. (2013). Model-free control. *International Journal of Control*, *86*(12), 2228–2252.

Godoy, J., Pérez, J., Onieva, E., Villagrá, J., Milanés, V., & Haber, R. (2015). A driverless vehicle demonstration on motorways and in urban environments. *Transport*, *30*(3), 253–263.

Guilloteau, Q., Robu, B., Join, C., Fliess, M., Rutten, É., & Richard, O. (2022). Model-free control for resource harvesting in computing grids. In *CCTA 2022*.

Han, Z., Xu, N., Chen, H., Huang, Y., & Zhao, B. (2018). Energy-efficient control of electric vehicles based on linear quadratic regulator and phase plane analysis. *Applied Energy*, *213*, 639–657.

Hossain, T., Habibullah, H., & Islam, R. (2022). Steering and speed control system design for autonomous vehicles by developing an optimal hybrid controller to track reference trajectory. *Machines*, *10*, 420.

Jiang, J., & Astolfi, A. (2018). Lateral control of an autonomous vehicle. *IEEE Transactions on Intelligent Vehicles*, *3*, 228–237.

K., V., Sheta, M. A., & Gumtapure, V. (2019). A comparative study of stanley, LQR and MPC controllers for path tracking application (ADAS/AD). In *2019 IEEE international conference on intelligent systems and green technology (ICISGT)* (pp. 67–674). IEEE.

Kebbati, Y., Ait-Oufroukh, N., Ichalal, D., & Vigneron, V. (2022). Lateral control for autonomous wheeled vehicles: A technical review. *Asian Journal of Control*.

Kebbati, Y., Puig, V., Ait-Oufroukh, N., Vigneron, V., & Ichalal, D. (2021). Optimized adaptive MPC for lateral control of autonomous vehicles. In *2021 9th international conference on control, mechatronics and automation (ICCMA)* (pp. 95–103). IEEE.

Lee, J., & Yim, S. (2023). Comparative study of path tracking controllers on low friction roads for autonomous vehicles. *Machines*, *11*, 403.

Lin, F., Chen, Y., Zhao, Y., & Wang, S. (2019). Path tracking of autonomous vehicle based on adaptive model predictive control. *International Journal of Advanced Robotic Systems*, *16*(5), Article 1729881419880089.

Liu, W., Hua, M., Deng, Z., Meng, Z., Huang, Y., Hu, C., Song, S., Gao, L., Liu, C., Shuai, B., Khajepour, A., Xiong, L., & Xia, X. (2023). A systematic survey of control techniques and applications in connected and automated vehicles. *CoRR*, abs/2303.05665.

Mata, S., Zubizarreta, A., & Pinto, C. (2019). Robust tube-based model predictive control for lateral path tracking. *IEEE Transactions on Intelligent Vehicles*, *4*(4), 569–577.

Menhour, L., d'Andréa Novel, B., Fliess, M., Gruyer, D., & Mounier, H. (2017). An efficient model-free setting for longitudinal and lateral vehicle control. Validation through the interconnected pro-SiVIC/RTMaps prototyping platform. *IEEE Transactions on Intelligent Transportation Systems*, *18*.

Moreno-Gonzalez, M., Artuñedo, A., Villagra, J., Join, C., & Fliess, M. (2022). Speed-adaptive model-free lateral control for automated cars. In *Joint IFAC conference: SSSC - TDS - LPVS, Montreal, IFAC 2022, Vol. 55* (pp. 84–89). 8th IFAC Symposium on System Structure and Control SSSC 2022.

Moreno-Gonzalez, M., Artuñedo, A., Villagra, J., Join, C., & Fliess, M. (2023). Speed-adaptive model-free path-tracking control for autonomous vehicles: Analysis and design. *Vehicles*, *5*(2), 698–717.

Pacejka, H. B., & Bakker, E. (1992). The magic formula tyre model. *Vehicle System Dynamics*, *21*, 1–18.

Park, M.-W., Lee, S.-W., & Han, W.-Y. (2014). Development of lateral control system for autonomous vehicle based on adaptive pure pursuit algorithm. In *2014 14th international conference on control, automation and systems (ICCAS 2014)* (pp. 1443–1447).

Pereira, G. C., Wahlberg, B., Pettersson, H., & Mårtensson, J. (2023). Adaptive reference aware MPC for lateral control of autonomous vehicles. *Control Engineering Practice*, *132*, Article 105403.

Peterson, M. T., Goel, T., & Gerdes, J. C. (2022). Exploiting linear structure for precision control of highly nonlinear vehicle dynamics. *IEEE Transactions on Intelligent Vehicles*, *8*(2), 1852–1862.

Rajamani, R. (2011). *Vehicle dynamics and control*. Springer Science & Business Media.

Rizk, H., Chaibet, A., & Kribèche, A. (2023). Model-based control and model-free control techniques for autonomous vehicles: A technical survey. *Applied Sciences*, *13*, 6700.

Samak, C. V., Samak, T. V., & Kandhasamy, S. (2021). Control strategies for autonomous vehicles. In *Autonomous driving and advanced driver-assistance systems (ADAS)* (pp. 70–140). Boca Raton: CRC Press, chapter 5.

Sorniotti, A., Barber, P., & Pinto, S. D. (2016). Path tracking for automated driving: A tutorial on control system formulations and ongoing research. In *Automated driving: Safer and more efficient future driving* (pp. 71–140). Springer International Publishing.

Stano, P., Montanaro, U., Tavernini, D., Tufo, M., Fiengo, G., Novella, L., & Sorniotti, A. (2022). Model predictive path tracking control for automated road vehicles: A review. *Annual Reviews in Control*.

Stano, P., Montanaro, U., Tavernini, D., Tufo, M., Fiengo, G., Novella, L., & Sorniotti, A. (2023). Model predictive path tracking control for automated road vehicles: A review. *Annual Reviews in Control*, *55*, 194–236.

Villagra, J. (2023). Interplay between decision and control. In *Decision-making techniques for autonomous vehicles* (pp. 193–213). Elsevier.

Villagra, J., & Herrero-Perez, D. (2012). A comparison of control techniques for robust docking maneuvers of an AGV. *IEEE Transactions on Control Systems Technology*, *20*(4), 1116–1123.

Villagra, J., Join, C., Haber, R., & Fliess, M. (2020). Model-free control for machine tools. In *21st IFAC world congress, IFAC 2020*.

Wang, Z., Sun, K., Ma, S., Sun, L., Gao, W., & Dong, Z. (2022). Improved linear quadratic regulator lateral path tracking approach based on a real-time updated algorithm with fuzzy control and cosine similarity for autonomous vehicles. *Electronics*, *11*(22), 3703.

Yakub, F., & Mori, Y. (2015). Comparative study of autonomous path-following vehicle control via model predictive control and linear quadratic control. In *Proceedings of the institution of mechanical engineers, Part D: Journal of automobile engineering, Vol. 229* (pp. 1695–1714). SAGE Publications Ltd.

Zainal, Z., Rahiman, W., & Baharom, M. (2017). Yaw rate and sideslip control using pid controller for double lane changing. *Journal of Telecommunication, Electronic and Computer Engineering (JTEC)*, *9*(3–7), 99103.

Zhao, P., Chen, J., Song, Y., Tao, X., Xu, T., & Mei, T. (2012). Design of a control system for an autonomous vehicle based on adaptive-PID. *International Journal of Advanced Robotic Systems*, *9*(2), 44.

Ziane, M. A., Pera, M., Join, C., Benne, M., Chabriat, J., Steiner, N. Y., & Damour, C. (2022). On-line implementation of model free controller for oxygen stoichiometry and pressure difference control of polymer electrolyte fuel cell. *International Journal of Hydrocarbon Engineering*.